\documentclass[aps,prb,twocolumn,superscriptaddress,floatfix]{revtex4}
\usepackage{graphicx}
\usepackage{psfrag}
\usepackage{color}
\usepackage{subfigure}
\usepackage{amsmath}
\definecolor{dred}{rgb}{0.7,0.0,0.0}
\allowdisplaybreaks 
 
\begin{document}

\title{Magnetic Phase Diagram of a Five-Orbital Hubbard Model in the  \\
Real-Space Hartree Fock Approximation Varying the Electronic Density}

\author{Qinlong Luo}

\author{Elbio Dagotto}

\affiliation{Department of Physics and Astronomy, University of
Tennessee, Knoxville, Tennessee 37996, USA} 
\affiliation{Materials Science
and Technology Division, Oak Ridge National Laboratory, Oak Ridge,
Tennessee 37831, USA}

\date{\today}

\begin{abstract}
Using the real-space Hartree Fock approximation, the magnetic phase diagram of a
five-orbital Hubbard model for the iron-based superconductors is studied
varying the electronic density $n$ in the range from 5 to 7 electrons per transition metal atom.
The Hubbard interaction $U$ is also varied, at a fixed Hund coupling $J/U=0.25$.
Several qualitative trends and a variety of competing magnetic states are observed. 
At $n$=5, a robust G-type antiferromagnetic
insulator is found, in agreement with experimental results for BaMn$_2$As$_2$. 
As $n$ increases away from 5, magnetic states with an increasing number of 
nearest-neighbors ferromagnetic links become energetically stable. This 
includes the well-known C-type antiferromagnetic state at $n$=6, 
the E-phase known to exist in FeTe, and also a variety of novel states not found yet experimentally, 
some of them involving blocks of ferromagnetically oriented spins. 
Regions of phase separation, as in Mn-oxides, have also been detected. 
Comparison with previous theoretical investigations indicate that 
these qualitative trends may be generic characteristics 
of phase diagrams of multiorbital Hubbard models.

\end{abstract}

\maketitle

\section{Introduction}

The study of iron-based high critical temperature superconductors
continues attracting the attention of the condensed matter community.\cite{johnston} 
Early theoretical investigations suggested a relatively simple picture of the magnetic and
superconducting properties as arising from weak-coupling Fermi surface nesting effects. However, recent experimental
and theoretical studies have unveiled a variety of compounds and chemical compositions that
display a more complex physics where intermediate-range electronic repulsion effects 
cannot be neglected.\cite{dai-review} In particular, there are materials with no Fermi surface nesting that 
nevertheless become superconducting, and there are compounds with a very large
magnetic moment in the ground state that do not fit into the weak coupling picture.\cite{RMP-2013} 
Moreover, at room temperature clear indications
of local magnetic moments exist,\cite{roomT} incompatible with weak coupling scenarios where
the formation of moments and the long-range order develop simultaneously upon cooling.

For these reasons, a more serious consideration of the effects of the 
Hubbard on-site repulsion $U$ and on-site Hund coupling $J$ is needed.
While this task is in principle difficult due to the scarcity of unbiased many-body techniques
that can handle a multiorbital Hubbard model, the use of mean-field approximations can at least unveil
qualitative tendencies in phase diagrams and the characteristics of the dominant states.
In fact, the Hartree and Hartree Fock approximations have been recently successfully used by our
group\cite{rong,luo} and others\cite{bascones104,bascones105} 
to study the dominant states in the presence of the $\sqrt{5} \times \sqrt{5}$
distribution of iron vacancies that exists in some selenides\cite{vacancies} 
and also for the case of two-leg ladder geometries.\cite{ladders} In all these cases,
the phase diagrams involve several different magnetic states and for this reason 
phase competition is anticipated to occur.

Also in more recent times, a novel avenue of research  motivated
by the iron-based superconductors has been expanding. 
It consists of replacing entirely Fe by another $3d$ transition
element such as Mn or Co. The average electronic population of these elements
in the new compounds
is different from that of iron, but the crystal structures are similar. Thus, as a first
approximation this chemical substitution
effectively amounts to exploring the effects 
of varying substantially the electronic density away from the original density of the iron-based materials. 
For example, in the case of the 100\% replacement of Fe by Mn,
the compound BaMn$_2$As$_2$ was found to develop a G-type antiferromagnetic (AFM) state with staggered
spin order, a N\'eel temperature of 625~K, and a magnetic moment
of 3.88$\mu_B$/Mn at low temperatures.\cite{BMA} The G-type AFM order is very robust, as recent investigations
of Ba$_{1-x}$K$_x$Mn$_2$As$_2$ have unveiled.\cite{persistence} This state 
emerges naturally from the
population $n$=5 at each Mn atom, namely 
one electron per $3d$ orbital.
In the other limit of full Co substitution for Fe, such as for the case of SrCo$_2$As$_2$, the material has
a complex Fermi surface and there are tendencies to magnetic order in the form of spin fluctuations
in the C-type channel,\cite{SCA} 
although $ab-initio$ calculations suggest that a ferromagnetic instability can also occur
(for a list of recent references see Ref.~\onlinecite{SCA}). 
Note that ferromagnetic tendencies have been reported 
for LaCoO$X$ ($X$=P,As) as well.\cite{LCOX}

These interesting recent studies motivate the model Hamiltonian
investigations reported here where the electronic density per transition metal atom, $n$, 
is allowed to vary over a wide range, centered
at the $n$=6 value corresponding to pnictides 
and selenides where the ground state is a C-type antiferromagnet. 
In previous efforts, the G-type AFM state at $n$=5 was already reported.\cite{bascones2,bascones3}
Other investigations assign a crucial role to the $n$=5 G-type AFM state
to understand the physics of the $n$=6 limit.\cite{imada}
In some studies the superconducting state
of pnictides is visualized as emerging from the $n$=5 G-type insulator\cite{capone} 
as opposed to being induced from the C-type
antiferromagnetic metal of $n$=6. All these previous efforts provide additional motivation for our studies. 
Thus, drastically altering the electronic concentration far away
from $n$=6 may lead to interesting perspectives to understand the pnictide and selenide superconductors.

The main result of this publication is the phase diagram
of a five-orbital Hubbard model in the real-space Hartree Fock approximation,
varying $U$ at fixed $J/U$ and,
more importantly, the electronic density from $n$=5 to 7. Three
main tendencies have been identified: {\it (i)} There are multiple 
magnetic states competing for space in the phase diagram. This is
indicative of a complex landscape of free energies. The results are compatible
with several states already unveiled experimentally for different
compounds,\cite{johnston,RMP-2013} and with other recent mean-field 
studies as well,\cite{bascones2,bascones3} but there are phases in the present theoretical
phase diagram that are novel and worth searching for experimentally. 
{\it (ii)} The general tendency in the evolution of the
magnetic states with increasing $n$ is to evolve from the G-AFM state at $n$=5
to states with more ferromagnetic links as $n$=7 is approached, particularly at robust $J/U$. 
{\it (iii)} There are regions in the phase diagram that present
the phenomenon of phase separation. This phenomenon was widely discussed before in
manganites,\cite{yunoki1,yunoki2,manganites,SSC} 
but it is only recently that this effect has been mentioned in the context
of the iron-based superconductors and their consequences are still unclear.
Our present conclusions are compatible with theoretical results
by other groups that also reported phase separation tendencies,\cite{bascones2,nori1,nori2} 
and also with previous investigations by our group that revealed 
the presence of stripes in some models.\cite{luo-stripes}

The organization of the results is the following. In Sec.~II, the model
and details of the calculations are explained. In Sec.~III, the main results
and phase diagram are presented. Sections IV and V include the results 
addressing the density-of-states and phase separation tendencies, respectively.
Finally, conclusions are presented in Sec.~VI.

\section{Model}
In this effort a five-orbital Hubbard
model will be used, with emphasis on the magnetic
states that are obtained by varying couplings and the electronic density $n$. 
Superconducting tendencies will not be investigated in the present
study. The model used is exclusively based on electrons that are located in 
the Fe $3d$ orbitals, 
widely believed to be the most relevant
orbitals at the Fermi surface for the pnictides and selenides.
Moreover, recent angle-resolved photoemission studies of BaCo$_2$As$_2$ compared with BaFe$_2$As$_2$ suggest that
a nearly rigid shift of the Fermi  level accounts for the complete substitution of Co for Fe,\cite{arpes} thus further
motivating our use of a single model with varying chemical potential to study a variety of materials.

The model includes a tight-binding term defined as
\begin{eqnarray}\label{E.H0k}
H_{\rm TB} = \sum_{<\mathbf{i,j}>} \sum_{\alpha,\beta,\sigma}
t^{\alpha\beta}_{\bf ij} (c^\dagger_{\mathbf{i},\alpha,\sigma}
c^{}_{\mathbf{j},\beta,\sigma} + h.c.),
\end{eqnarray}
\noindent where $c^\dagger_{\mathbf{i},\alpha,\sigma}$ creates an electron with spin
$\sigma$ at the orbital $\alpha$ of the transition metal 
site $\mathbf{i}$ (a square lattice is used), 
and $t^{\alpha\beta}_{\bf ij}$ refers to the tunneling amplitude of an electron hopping
from orbital $\alpha$ at site $\mathbf{i}$ to orbital $\beta$ at site
$\mathbf{j}$. The Coulombic interacting portion of the five-orbital Hubbard 
Hamiltonian is standard and given by:
\begin{equation}\begin{split}  \label{eq:Hcoul}
  H_{\rm int}& =
  U\sum_{{\bf i},\alpha}n_{{\bf i},\alpha,\uparrow}n_{{\bf i},
    \alpha,\downarrow}
  +(U'-J/2)\sum_{{\bf i},
    \alpha < \beta}n_{{\bf i},\alpha}n_{{\bf i},\beta}\\
  &\quad -2J\sum_{{\bf i},\alpha < \beta}{\bf S}_{\bf{i},\alpha}\cdot{\bf S}_{\bf{i},\beta}\\
  &\quad +J\sum_{{\bf i},\alpha < \beta}(d^{\dagger}_{{\bf i},\alpha,\uparrow}
  d^{\dagger}_{{\bf i},\alpha,\downarrow}d^{\phantom{\dagger}}_{{\bf i},\beta,\downarrow}
  d^{\phantom{\dagger}}_{{\bf i},\beta,\uparrow}+h.c.),
\end{split}\end{equation}

\noindent where $\alpha,\beta$ denote the five 
$3d$ orbitals with a label convention defined in Table III below, 
${\bf S}_{{\bf i},\alpha}$ ($n_{{\bf i},\alpha}$) is 
the spin (electronic density) of orbital $\alpha$ at site
${\bf i}$, and the relation $U'=U-2J$ between the Kanamori parameters
has been used.
The first two terms give
the energy cost of having two electrons located in  the same orbital or in
different orbitals, both at the same site, respectively. The third term is the Hund's
coupling that favors the ferromagnetic (FM) alignment of the spins in
different orbitals at the same lattice site. 
The ``pair-hopping'' is the fourth term and its
coupling is equal to $J$ by symmetry.

\begin{figure}[thbp]
\begin{center}
\includegraphics[width=9.0cm,clip,angle=0]{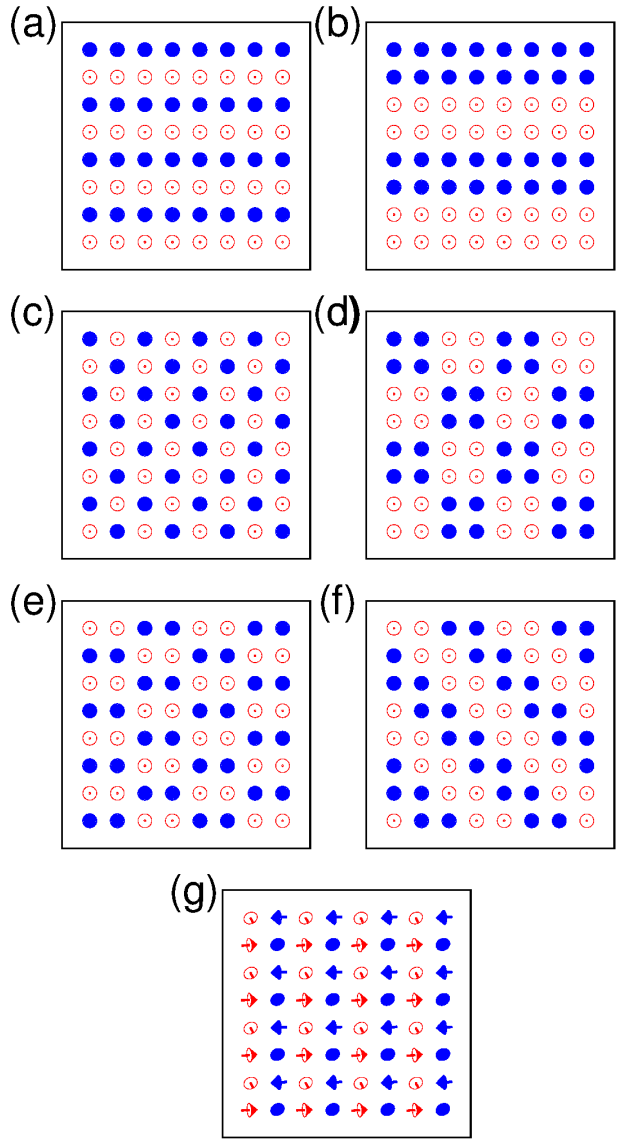}
\caption{ Magnetic states observed in the phase diagram of the
five-orbital Hubbard model used in this study, treated in the HF approximation. 
These magnetic states are named as: 
(a) C, (b) DC, (c) G, (d) Block, (e) GC, (f) E, and (g) Flux. }
\label{fig-spins}
\end{center}
\end{figure}

With regards to the tight-binding parameters, the set of hoppings used in the present effort 
is taken from Ref.~\onlinecite{Graser08}, 
which provides a Fermi surface that compares well with experiments and band structure calculations for the pnictides. 
The overall conclusions of our study are sufficiently generic that they are likely to be valid even if other
set of hoppings are used, although certainly the details and actual critical couplings will change from set to set. 
The actual hoppings employed here are provided in Table III in the Appendix.
The approximate bandwidth $W$ of the tight-binding hopping term is 4.7 eV, 
and the ratio $U/W$ should be used to judge whether 
the phases of interest are or not, e.g.,  in the strong coupling regime where $U/W \sim 1$. A ratio
$U/W \sim 0.5$ is more typical for the location of the experimentally relevant phases 
based on previous Hartree Fock investigations,\cite{dai-review,luo} 
signaling an intermediate coupling regime. However, note that
the quantum fluctuations not considered in mean-field studies will tend to increase the critical values of $U/W$.

To study the ground state properties of the multiorbital Hubbard model, 
the Hartree Fock (HF) approximation will 
be applied to the Coulombic interaction. The HF Hamiltonian is solved by minimizing the 
energy via a numerical real-space 
self-consistent iterative process that was widely discussed in previous
efforts.\cite{vacancies,ladders,luo-stripes} 
All the HF expectation values are initially 
assumed independent from site to site, which allows the system to select spontaneously the state 
that minimizes the HF energy, reducing the bias into 
the calculations. In the self-consistent iterative process, initially all the HF expectation values are set to random numbers, 
physically corresponding to random initial states. The iterative process converges to states that resemble uniformly ordered
states, albeit still with some deviations that are difficult to remove in the (slow) iterative process.
Inspired by the results obtained with random starts, then fully ordered starting 
configurations are also used as starting points for comparison. 
At the end, the ground states are selected by comparing the final energies after convergence.
In Fig.~\ref{fig-spins} the reader can find the set of relevant states 
that appeared spontaneously in the real-space energy minimization used in the present effort. 
All the numerical results are obtained using a real-space 8$\times$8 square 
lattice with periodic boundary conditions.
The criteria of convergence is set so that the changes of the individual 
HF expectation values are less than $10^{-4}$. Under this criteria, 
the typical number of iterations is from $500$ to $1000$ if random initial states are used, 
and from $50$ to $200$ if the starting configurations
correspond to ordered states as those in Fig.~\ref{fig-spins}.

\section{Main Results}

\subsection{Phase Diagram}

\begin{figure}[thbp]
\begin{center}
\includegraphics[width=9.1cm,clip,angle=0]{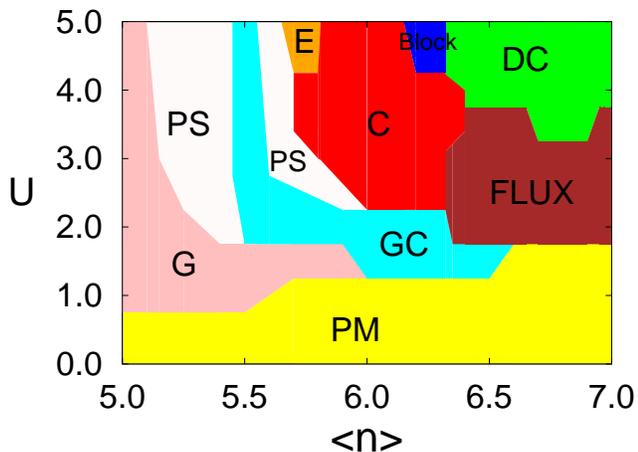}
\caption{Phase diagram of the five-orbital Hubbard model varying the on-site same-orbital repulsion $U$ and
the electronic density $n$ (number of electrons per transition metal site). 
The Hund coupling was fixed to $J/U= 0.25$. The notation for the many states was explained
in Fig.~1.
Light pink areas correspond to ``Phase Separation'' (PS) regions where the energy vs. $n$ curves have a negative
curvature (as described later in the manuscript). In practice, at least a vestige of magnetic order is typically found in the
numerical process even for very small values of $U$. However, previous experience indicates that this is likely a 
``Paramagnetic'' (PM) state since it is smoothly connected to the $U=0$ limit. Thus, in practice the PM state is
defined as the state where the order parameter $m$, of any kind, is smaller than a cutoff chosen as $4\%$ of the saturated value 
for the same state at other densities or couplings. Since the order parameters often raise steeply at the critical $U$ 
that separates the PM from magnetic states, then selecting
other cutoffs give similar results. Note also that the bandwidth $W$ of the hopping term is 4.7 eV.}
\label{PhaseDiag}
\end{center}
\end{figure}

The effort described in this publication was computationally intense, since there were two parameters to change ($U$ and $n$;
$J/U$ was fixed to 0.25, a value considered realistic from previous investigations\cite{luo}) 
and the real-space HF process is typically characterized
by a slow convergence in the iterative process. The main result of this study is 
summarized in the HF phase diagram of the five-orbital Hubbard model, varying the on-site 
coupling $U$ and electron density $n$, shown in Fig.~\ref{PhaseDiag}. 

Let us now describe in detail the results. Starting at $n=5$, i.e. 5 electrons 
for the five $3d$ transition metal orbitals, the state has a strong
tendency to form a G-type AFM state. This is to be expected given the electronic population, and this result is in
excellent agreement with experiments\cite{BMA} for BaMn$_2$As$_2$ and 
with previous theoretical efforts.\cite{bascones2,imada,capone}
The robustness of the G-AFM state suggests that using other hoppings amplitudes, such as those
more quantitatively adequate to describe BaMn$_2$As$_2$, will likely lead to similar conclusions.

The G-AFM state has individual spins that are antiferromagnetically
coupled to their four neighbors. As $n$ increases, growing tendencies toward developing
more ferromagnetic links are observed. In fact, the novel ``GC'' state (see Fig.~1) is stabilized next when 
increasing $n$ away from 5, and this state has three AFM links and one FM link.
This state can be considered as a combination of the G-AFM and C-AFM states, thus the notation GC. 
Its dominant wave vector is ($\pi/2$,$\pi$), and the state breaks rotational invariance
between the two axes $x$ and $y$, as the C-AFM states does, 
but also has a staggered ordered as the G-AFM state does, although involving 2$\times$1 blocks.
Thus, with hindsight it is not 
surprising to find this GC state stabilized in between
the G and C states.
A somewhat surprising result is that the area of stability of the GC state also includes
a region of weak $U$ coupling at $n=6$ where it is widely believed that the C-type AFM
state should dominate. This C-AFM state indeed is stable increasing $U$ but not at very weak
coupling. 
Considering that recent Monte Carlo computational studies including lattice distortions 
and using three orbitals in the context of a spin-fermion model do favor 
the C-AFM state,\cite{shuhuaPRL1,shuhuaPRL2}
then probably the absence of lattice degrees of freedom in the present effort may lead to
a spurious larger region of stability of the GC state that includes portions of the $n=6$ axis. 
Thus, readers should be
warned that the region of true stability of the GC-AFM state may be smaller than the HF
approximation suggests, particularly after lattice effects and quantum fluctuations are incorporated.
In general, only qualitative trends are expected to be robust in the
present study but not detailed quantitative aspects. The prediction arising from this effort 
is that it would not be surprising to find the GC state stabilized in materials 
where the relevant electronic density is approximately $n$=5.5.

As already mentioned, centered at $n=6$ and for intermediate and large $U$ the C-AFM state
is stabilized, in agreement with many experiments and several 
other theoretical studies.\cite{johnston,dai-review,RMP-2013} 
Since this state has been widely discussed before in many contexts, there is no need
to repeat those discussions and the focus here now shifts to values of $n$ larger than 6.
In this regime, several exotic states are stabilized in the HF approximation. 
One of these novel states is the ``Flux'' state, shown in Fig.~1(g). Note that this state is not collinear.
A similar state has
been discussed before in the context of two-orbital Hubbard models,\cite{lorenzana} and in small regions 
of the phase diagram of a five-orbital
Hubbard model defined on two-leg ladders.\cite{ladders} To our knowledge 
this Flux state has not appeared in previous studies when using  
two-dimensional geometries and five-orbital models, and it has not been observed experimentally yet. 

Another exotic and novel state stabilized 
at $n$ larger than 6 is the double-C, ``DC'', state shown in Fig.~1(b). 
The notation double C is in reference 
to the doubled period in one direction with respect to the well-known C-state.
This DC state has a spin structure factor peaked at (0,$\pi/2$) 
or ($\pi/2$,0) depending on the lattice instabilities 
that may appear in a real system. This DC state is representative of
the previously-mentioned growing ferromagnetic tendencies 
with increasing $n$ since 
each spin has three (one) ferromagnetically (antiferromagnetically)
aligned neighbors. It is conceivable that with further
increasing $n$ and/or $U$ and $J$, a fully ferromagnetic state can be stabilized, 
as already observed in previous HF approximation studies in other contexts 
such as ladders and with iron vacancies.\cite{ladders,vacancies} 
Note also that from
our results near $n$=7 (Fig.~2) there are no indications 
that the C-type AFM state can become stable
at such large electronic densities, at least at the level of ground states. Thus, the recent inelastic 
neutron scattering results\cite{SCA} for SrCo$_2$As$_2$ reporting C-type fluctuations 
remain paradoxical, and deserve further studies.

In addition to the dominant G, GC, C, Flux, and DC states, there are two small regions where two
exotic states, the E and Block states, are stabilized. These states need a robust $U$ to become 
stable (i.e. $U/W \sim 1$ is needed for their stability)
and they have been mentioned in other contexts before. For instance, the Block-AFM state is made of
2$\times$2 FM blocks that are coupled antiferromagnetically. This state was proposed to be the ground
state of KFe$_2$Se$_2$ in previous theoretical investigations.\cite{dong-hu} 
A similar ``Block'' structure has been unveiled experimentally and theoretically 
in materials with iron vacancies\cite{RMP-2013,vacancies} 
and also in selenides with two-leg ladder geometries.\cite{ladders} 
These Block states have individual spins with two antiferromagnetic links
and two ferromagnetic links, thus their location next to the C-AFM state is reasonable since they share
this same property. This line of reasoning is mainly of relevance 
for discussions involving localized spins, as they occur
at robust $U$. It is gratifying that the Block-AFM appears spontaneously in our calculations without
the need of introducing lattice distortions.

The other exotic state stabilized in a small region at robust $U/W$
is the ``E'' state shown in Fig.~1(f). This state has a peak in the spin structure
factor located at ($\pi/2$,$\pi/2$), which is 
compatible with experimental neutron scattering results\cite{Ephase} 
for FeTe. 
Historically, the E phase was reported
initially in investigations of manganites.\cite{hotta} 
The existence of the E phase is also compatible with more recent theoretical studies 
that used the spin-fermion model, involving a mixture of localized and 
itinerant degrees of freedom with two active orbitals.\cite{weiku} 
The E state was also reported by another group in previous investigations of
a five-orbital Hubbard model, using momentum-space 
mean-field and Heisenberg techniques, and a different set of hopping amplitudes.\cite{bascones2} 
Note that in the previous publication Ref.~\onlinecite{bascones2} 
the E-state is actually called the DS-state. Here, the historical notation that
started with the manganites is used and the state is called E.
Note also that recent investigations suggest ferro-orbital
order and a bond-order wave in Fe$_{1.09}$Te in the regime of the E-phase,\cite{fobes}
implying that the region where the E-state is here reported to be stable 
should deserve further more detailed studies.

In summary, the four states G, C, E, and Block have been observed 
before in different materials of the
family of iron-based superconductors and in other theoretical studies, 
while the possible stability of the three states GC, Flux, and DC 
are original predictions of the present study.
Note that the mean-field approximation 
used here tends to exaggerate the presence of magnetic order.
While the predictions are expected to be reasonable at special density fractions such as $n$=5, 5.5, 6, ...,
the phase diagram unveiled here at intermediate values of $n$ 
is at best indicative of qualitative tendencies
that may exist, perhaps, only in the form of short-range correlations. 
Also note that superconducting states have not
been proposed in this mean-field study, so the focus is only on 
magnetic order (and its concomitant orbital order, as described below). 

\subsection{Magnetic Order Parameters}

In Table I, characteristic magnetic moments of the seven phases found in Fig.~2
are provided at representative couplings and densities. The values shown tend 
to indicate a robust magnetic moment. However, 
in the phases that are in contact with the weak coupling PM state in
the phase diagram (i.e. the G, GC, and Flux states), there is a region of rapid
change in the value of the magnetic moment when magnetism develops, as shown
in Figs.~3 and 4. Thus, values of the magnetic moments weaker than those
in Table I are also possible for some of the phases.

\begin{table}[thpd]
\begin{ruledtabular}
{\normalsize 
\begin{tabular}{|c|c|c|c|c|c|c|}
 & $xz$ & $yz$ & $x^2-y^2$ & $xy$ & $z^2$ & total \\
\hline
\hline
C      & 0.9235 & 0.5426 & 0.5678 & 0.9499 & 0.7451 & 3.7289 \\
\hline
Flux$^*$ & 0.6094 & 0.6735 & 0.4812 & 0.8372 & 0.5693 & 3.1692 \\
\hline
G      & 0.9475 & 0.9475 & 0.9242 & 0.9609 & 0.9682 & 4.7481 \\
\hline
GC      & 0.9362 & 0.7853 & 0.7027 & 0.9540 & 0.8625 & 4.2407 \\
\hline
E$^*$  & 0.8589 & 0.8602 & 0.6063 & 0.9843 & 0.8702 & 4.1799 \\
\hline
Block      & 0.8296 & 0.8296 & 0.6559 & 0.9573 & 0.3944 & 3.6667 \\
\hline
DC     & 0.7632 & 0.6043 & 0.5470 & 0.9102 & 0.3611 & 3.1858 \\
\end{tabular}
}
\end{ruledtabular}
\caption{\label{tableM} 
Magnetic moments of the seven competing states 
at selected couplings and densities. The details are as follows: 
C-state ($U$=3.0, $n$=6.0);
Flux-state ($U$=3.0, $n$=6.5);
G-state ($U$=3.0, $n$=5.0);
GC-state ($U$=3.0, $n$=5.5);
E-state ($U$=5.0, $n$=5.75);
Block-state ($U$=5.0, $n$=6.25);
DC-state ($U$=5.0, $n$=6.75). 
The phases with $\ast$ indicates that the magnetic moment is not the same at each site. 
Typically, there are four sites that repeat themselves in most of the cases, 
but sometimes the periodicity involves two sites or eight sites.
The numbers used for these states in the present table are their average values.
}
\end{table}

In Fig.~3, the order parameter at $n=6$ is explicitly shown, varying $U$. While the C-AFM state 
that is stabilized at intermediate and large $U$ is to be expected, the presence of the 
GC-AFM state in the weak coupling regime is a surprise, as already discussed.
In view of the many approximations involved 
in arriving to this state, it would be premature to claim that the GC-state
should be stable in portions of the phase diagram corresponding 
to the Fe-based compounds, but its presence in the phase diagram
can be considered as indicative of a competition between many magnetic states. 
In practice, other degrees of freedom, such
as the lattice, are probably crucial in deciding which state is the most stable in an actual compound.

\begin{figure}[thbp]
\begin{center}
\includegraphics[width=9.0cm,clip,angle=0]{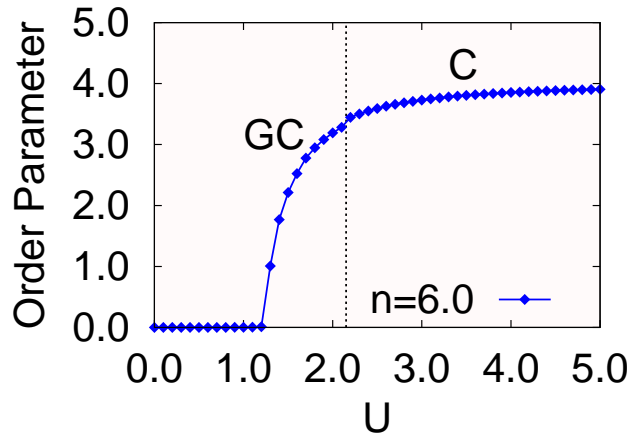}
\caption{ Hartree Fock order parameters (Bohr magneton units) vs. $U$ at density $n=6.0$ and $J/U=0.25$.
The magnetic states GC and C have been presented in Fig.~1. The bandwidth $W$ is 4.7~eV.
}
\label{mU1}
\end{center}
\end{figure}

Similar results were obtained at other electronic densities, as shown in Fig.~4.
At $n=5$, the G-AFM state is clearly dominant, with an order parameter (in units of the Bohr magneton) 
that tends to the maximum value 5 as $U$ grows.
At the other electronic densities shown, there
is always phase competition between two or three states, and this phase competition may preclude
the order parameters from reaching their maximum value, at least in the range studied.
The transitions between different magnetic states are of first order but the jumps in the
order parameters tend to be rather small and in some cases the curves look almost continuous.

\begin{figure}[thbp]
\begin{center}
\includegraphics[width=9.0cm,clip,angle=0]{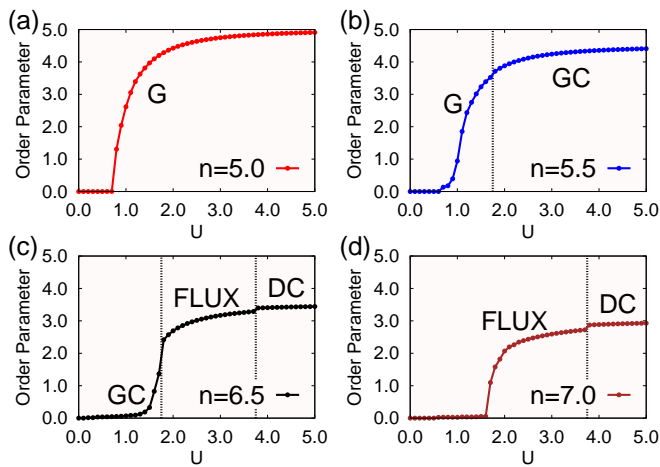}
\caption{ Hartree Fock order parameters (Bohr magneton units) 
vs. $U$ at $J/U=0.25$ and several electronic densities: 
(a) $n=5.0$; (b) $n=5.5$; (c) $n=6.5$; (d) $n=7.0$. All the states indicated 
are shown explicitly in Fig.~1. 
}
\label{mU2}
\end{center}
\end{figure}

\subsection{Orbital Composition}

The orbital compositions of the seven states unveiled in the phase diagram of Fig.~2 are given in Table II.
From the perspective of these occupations, the G-AFM state has clear indications of being an insulator
since all the five orbitals are approximately equally populated with one electron per orbital.
On the other hand, most of the orbitals of the other six states have 
a population substantially different from one, potentially 
giving rise to a metallic state (perhaps with coexisting itinerant and localized degrees of freedom).
However, the Block-AFM state should be 
insulating due to the peculiar spin geometry of the state that renders difficult
for electrons to transition from block to block while keeping the same spin orientation.

\begin{table}[thpd]
\begin{ruledtabular}
{\normalsize 
\begin{tabular}{|c|c|c|c|c|c|c|}
 & $xz$ & $yz$ & $x^2-y^2$ & $xy$ & $z^2$ & total \\
\hline
\hline
C      & 1.0048 & 1.3911 & 1.3659 & 1.0099 & 1.2281 & 6.0 \\
\hline
Flux$^*$ & 1.3083 & 1.2435 & 1.4462 & 1.1108 & 1.3912 & 6.5 \\
\hline
G      & 0.9998 & 0.9998 & 1.0029 & 0.9965 & 1.0009 & 5.0 \\
\hline
GC      & 1.0017 & 1.1606 & 1.2269 & 1.0025 & 1.1083 & 5.5 \\
\hline
E$^*$  & 1.1249 & 1.1236 & 1.3758 & 1.0046 & 1.1212 & 5.75 \\
\hline
Block      & 1.1509 & 1.1509 & 1.3244 & 1.0297 & 1.5940 & 6.25 \\
\hline
DC     & 1.2183 & 1.3796 & 1.4397 & 1.0807 & 1.6317 & 6.75 \\
\end{tabular}
}
\end{ruledtabular}
\caption{\label{tableN} 
Orbital compositions of the seven competing states 
at selected couplings and densities. The details are as follows: 
C-state ($U$=3.0, $n$=6.0);
Flux-state ($U$=3.0, $n$=6.5);
G-state ($U$=3.0, $n$=5.0);
GC-state ($U$=3.0, $n$=5.5);
E-state ($U$=5.0, $n$=5.75);
Block-state ($U$=5.0, $n$=6.25);
DC-state ($U$=5.0, $n$=6.75). 
Similarly as in Table I, the phases with $\ast$ indicates that the orbital population is not the same at each site. 
Typically, there are four sites that repeat themselves in most of the cases, 
but sometimes the periodicity involves two sites or eight sites.
The numbers used for these states in the present table are their average values.
}
\end{table}

\section{Density of States}

To investigate the metallic vs. insulating characteristics of the states presented in the
phase diagram, the density of states has been analyzed. The results are shown in Fig.~5. The
situation for the G-AFM state is clear: the state is an insulator with 
a robust gap. The Block-AFM state involving spin blocks is also insulating, as discussed above. 
This can be understood since in the Block-state 
there are no paths from one extreme to the 
other of the crystal with spins displaying the same spin orientation.

The C-AFM state is metallic, in agreement with previous calculations,\cite{luo} and the DC state is also
metallic. This is reasonable since C and DC only differ in the periodicity 
along the $y$ direction (strictly speaking,
for the 8$\times$8 cluster there is a tiny gap in the DOS for the DC-state 
but this is likely caused by finite-size effects).
The E-phase also displays a small gap, but it is difficult to say whether it will become insulating 
or metallic in the bulk limit. Finally, the Flux state appears to be clearly metallic, while the GC-AFM state is 
insulating. The latter 
has this property because it is formed by isolated 2$\times$1 spin blocks, 
qualitative similar to the characteristics that led to 
the insulating nature
of the Block-AFM state made of isolated 2$\times$2 spin blocks.

\begin{figure}[thbp]
\begin{center}
\includegraphics[width=9.0cm,clip,angle=0]{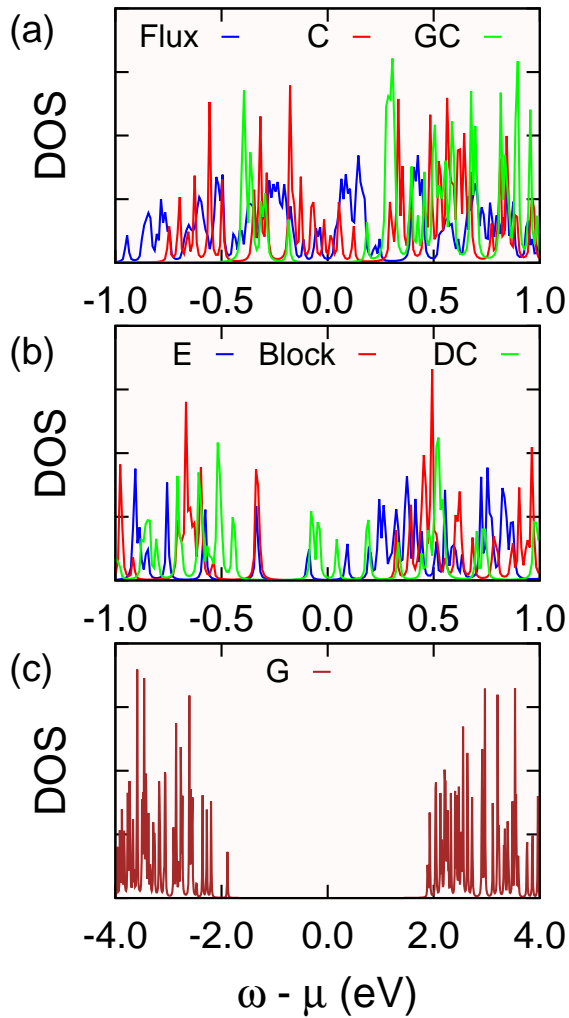}
\caption{ Density of States (DOS) at representative values of couplings and densities, corresponding to the seven
magnetic states that appear in the phase diagram of Fig.~2. (a) Flux-state ($U$=3.0, $n$=6.5);
C-state ($U$=3.0, $n$=6.0);
GC-state ($U$=3.0, $n$=5.5);
(b) 
E-state ($U$=5.0, $n$=5.75);
Block-state ($U$=5.0, $n$=6.25);
DC-state ($U$=5.0, $n$=6.75);
(c) G-state ($U$=3.0, $n$=5.0).
}
\label{DOS}
\end{center}
\end{figure}

\section{Phase Separation}

\begin{figure}[thbp]
\begin{center}
\includegraphics[width=9.0cm,clip,angle=0]{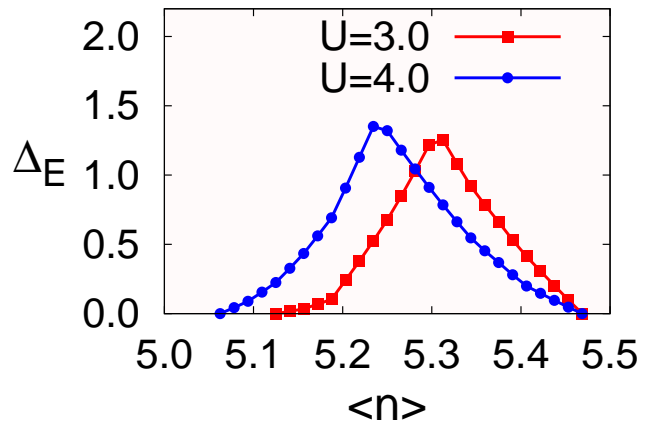}
\caption{ Plots of $\Delta E(n)$ vs $n$ showing the existence of negative curvature, namely phase
separation. The results were obtained for $U=4.0$ and $U=3.0$, $J/U=0.25$, and in the range of densities
indicated.
Here $\Delta E(n) = E(n) - E_0(n)$, where $E(n)$ is the actual ground 
state energy at electronic density $n$ and $E_0(n)$ is a straight line that 
joins the energies of the two densities at the boundaries of the PS regions. }
\label{PS}
\end{center}
\end{figure}

The phase diagram shown in Fig.~\ref{PhaseDiag} contain regions of phase separation (PS). 
The conclusion that there are unstable regions with these characteristics in the phase diagram
was based on the study of the curvature of the $E(n)$ vs. $n$ curves, where $E(n)$ is the
ground state energy at the electronic density $n$. Phase separation in multiorbital systems
occurs in other contexts, such as in double exchange models 
for manganites,\cite{yunoki1,yunoki2,manganites,SSC} thus it is not
unexpected to find the same phenomenon 
in the five-orbital Hubbard model as well.
In order to visualize the presence of regions with negative curvature 
in the $E(n)$ vs. $n$ curves
it is better to introduce $\Delta E(n) = E(n) - E_0(n)$, 
where $E_0(n)$ is a straight line that joins the energies at the 
boundary densities of the PS region. Therefore, $\Delta E(n)$ should be positive 
if $E(n)$ vs. $n$ has a negative
curvature. Some representative results are shown in Fig.~\ref{PS}, where indeed it is clear that PS exist in 
the regimes of parameter space 
corresponding to those curves. 

The two regions in which the PS state separates are in principle
macroscopic in size. However,
previous experience with Mn-oxides\cite{manganites} suggest that
once other interactions are included, particularly the long-range 
portion of the Coulomb repulsion between electrons,
the PS regions become unstable. 
This macroscopic separation is replaced instead by complex 
states that are mixtures, at the nanometer length
scale, of the two phases at the boundaries
of the PS portions of the phase diagram. 
In this regime, nonlinear responses to external fields could be expected.\cite{manganites}

Note that phase separation was also observed in previous studies of multiorbital
Hubbard models, employing related momentum-space mean-field
and Heisenberg mean-field techniques, and a different set of hopping amplitudes.\cite{bascones2} 
In particular, the 
PS regions of Ref.~\onlinecite{bascones2} also involve the G and C states as in our results, although
in our case the GC state (not included in the study of Ref.~\onlinecite{bascones2}) also plays an important role.
Although the agreement is not quantitative, the similarities 
of both studies suggest that PS must be considered
when phase diagrams of multiorbital Hubbard models are constructed. As mentioned before, 
the presence of PS was also reported
in recent related calculations that employed a mean-field approximation to a model with weakly coupled
electrons having an electron- and a hole-band with imperfect nesting.\cite{nori1,nori2} 
The qualitative agreement with these previous results suggest once again 
that the PS tendency may be generic and should be considered into future studies, and even in the
interpretation of some experiments. 

\section{Conclusions}

\begin{table*}[thpd]
\centering
\begin{ruledtabular}
{ \normalsize
 \begin{tabular}{|c|c|c|c|c|c|c|c||c|}
$t^{mn}_i$ & $i=x$ & $i=y$ & $i=xy$ & $i=xx$ & $i=xxy$ & $i=xyy$ & $i=xxyy$ & $\epsilon_{mn}$ \\
\hline
  $mn=11$  & -0.14 & -0.4  & 0.28   & 0.02   & -0.035  & 0.005   & 0.035    &    0.13   \\
  $mn=33$  & 0.35  &       & -0.105 & -0.02  &         &         &          &   -0.22   \\
  $mn=44$  & 0.23  &       &  0.15  & -0.03  & -0.03   &         & -0.03    &    0.3    \\
  $mn=55$  & -0.1  &       &        & -0.04  &  0.02   &         & -0.01    &   -0.211  \\
  $mn=12$  &       &       &  0.05  &        & -0.015  &         & 0.035    & \\
  $mn=13$  &-0.354 &       & 0.099  &        & 0.021   &         &          &\\
  $mn=14$  & 0.339 &       & 0.014  &        & 0.028   &         &          &\\
  $mn=15$  &-0.198 &       & -0.085 &        &         &         & -0.014   &\\
  $mn=34$  &       &       &       &       & -0.01  &        &        &\\
  $mn=35$  & -0.3  &       &       &       & -0.02  &        &        &\\
  $mn=45$  &       &       & -0.15 &       &        &        &  0.01  &  \\
 \end{tabular}
\caption{Hopping amplitudes for the tight-binding portion of the five-orbital Hubbard model
  used in this study. Here $m$ and $n$ label the Fe $3d$ 
orbitals as follows: $1=xz$, $2=yz$, $3=x^2-y^2$, $4=xy$, $5=3z^2$-$r^2$, 
  and $i$ labels the hopping directions. $\epsilon_{mn}$ in the last column is the on-site energy. 
  The explicit form of the tight-binding Hamiltonian can be found in Ref.~\onlinecite{Graser08}.
  The overall energy unit is electron volts.
  }
  \label{tab:hopp5}} 
\end{ruledtabular}
\end{table*}

The phase diagram of a 
five-orbital Hubbard model has been presented in this publication, working
at a fixed Hund coupling $J/U=0.25$, 
varying the Hubbard repulsion $U$ and the electronic density $n$
in the range from 5 to 7, and employing the real-space 
Hartree Fock approximation as the many-body technique. 
While our results cannot be considered quantitatively 
accurate, due to the intrinsic deficiencies of
mean-field approximations, qualitative trends 
appear reasonable and moreover they are in good agreement
with other independent theoretical investigations. These 
trends include the presence of many competing
magnetic states (superconducting states were 
not studied here), suggesting a rich free energy landscape
with several local minima. Perhaps not surprisingly based 
on previous studies on colossal magnetoresistive 
Mn-oxides, this rich landscape may lead to regions 
of phase separation where 
complex states involving a nanometer-scale mixture of the 
competing phases could be stabilized. 
In addition, there is a clear tendency to evolve from fully antiferromagnetic states at $n$=5
to states with an increasing number of ferromagnetic links as $n$ grows. 

Several of the states that 
spontaneously appeared in our phase diagram are known to exist in experiments, either for layered materials
or in other geometries such as with regularly spaced Fe vacancies or in two-leg ladders. These states
are the G-, E-, and C-type antiferromagnets, and also 
the 2$\times$2 Block state. In addition,  three novel states
have been found in our study: the GC, Flux, and DC states. 
Experimental efforts should be devoted
to the search for these states in actual compounds. The similarity
of our results with the conclusions of other theoretical efforts give us
confidence that the trends studied here are robust and characteristics of multiorbital Hubbard models
in general, in the range of densities from 5 to 7 
electrons per transition metal atom. 

\section{Acknowledgment} The work of Q.L. was supported by 
the U.S. DOE, Office of Basic Energy Sciences, Materials
Sciences and Engineering Division.
The work of E.D. for this project was supported by
the National Science Foundation under Grant No. DMR-1104386. \\

\section{Appendix}

For completeness, the hopping amplitudes used in this study are given in Table~III.




\end{document}